\documentclass[aps,prd,twocolumn,groupedaddress,nofootinbib]{revtex4-1}

\pdfoutput=1

\usepackage[english]{babel}
\usepackage[latin1]{inputenc}
\usepackage{graphicx}
\usepackage{amsmath}
\usepackage{color}

\begin{document}

\title{Comment on ``A fractal LTB model cannot explain Dark Energy''}
%
%
\author{L. Cosmai} \email{leonardo.cosmai@ba.infn.it}
\affiliation{Istituto Nazionale di Fisica Nucleare, Sezione di Bari, via G. Amendola 173, 70126, Bari, Italy}

\author{G. Fanizza} \email{gfanizza@fc.ul.pt}
\affiliation{Instituto de Astrof\'isica e Ci\^encias do Espa\c{c}o, Faculdade de Ci\^encias da Universidade de Lisboa, Edificio C8, Campo Grande, P-1749-016, Lisbon, Portugal}

\author{F. Sylos Labini}\email{sylos@cref.it}
\affiliation{Centro   Ricerche Enrico Fermi,  I-00184, Roma, Italia}
\affiliation{INFN Unit\'a Roma 1, Dipartimento di Fisica, Universit\'a di
  Roma Sapienza, I-00185 Roma, Italia }

\author{L. Pietronero}
\email{luciano.pietronero@roma1.infn.it}
\affiliation{Enrico Fermi Research Center, Via Panisperna 89a, 00184 Roma, Italy}

\author{L. Tedesco}
\email{luigi.tedesco@ba.infn.it}
\affiliation{Istituto Nazionale di Fisica Nucleare, Sezione di Bari, Bari, Italy}
\affiliation{Dipartimento di Fisica, Universit\`a di Bari, Via G. Amendola 173, 70126, Bari, Italy}

\date{\today}

\begin{abstract}
We reply to the criticisms moved in \cite{Pasten:2022ggz} against our results presented in \cite{Cosmai:2018nvx}. In particular, we show that our fractal model has none of the problems claimed in \cite{Pasten:2022ggz}. The latters can be addressed to the overlooked nonlinear behaviour of the Einstein's equations.
\end{abstract}


\maketitle 
In a recent preprint, Past\'en et {\it al.} \cite{Pasten:2022ggz} have claimed that the fractal Lema\^itre-Tolman-Bondi (LTB) solution proposed by Cosmai et al. \cite{Cosmai:2018nvx} is not a viable fit for the large scale Hubble-Lema\^itre diagram. In fact in \cite{Cosmai:2018nvx} we have shown that a fractal distribution of matter is able to fit the UNION 2 data \cite{Amanullah:2010vv} and may provide then a possible explanation for the (apparent) acceleration in the late time Universe.

Past\'en et {\it al} argue that the solution derived in \cite{Cosmai:2018nvx} is a simple redefinition of the radial coordinate in the Friedmann-Lema\^itre-Robertson-Walker (FLRW) metric in a Universe filled with dark matter. As that, hence, it is just the Einstein-deSitter model: this statement is incorrect, as we are going to demonstrate.

In fact, the authors of \cite{Pasten:2022ggz} claim that, following the treatment of \cite{Cosmai:2018nvx}, the solution for the proposed LTB model is given by their Eq.~(28). This equation is stated as the correct representation of the LTB model in their paper and is given by
\begin{equation}
R(t,r)=\left( \frac{9 M(r)}{2} \right)^{1/3} t^{2/3}\,,
\label{eq:wrong}
\end{equation}
where $M(r)$ is the mass content in a sphere of radius $r$ which scales as $r^D$ for fractal distribution of matter, where $D$ is the fractal dimension. 
What the authors in \cite{Pasten:2022ggz} argue is that this solution is just an overlooked gauge fixing which leads to Eq.~(6) of \cite{Cosmai:2018nvx}, which is (in  \cite{Cosmai:2018nvx} $A(t,r)$ is $R(t,r)$)
\begin{equation}
A(t,r)=A_0(r) \left[ 1+\frac{3}{2}\,\sqrt{\frac{2GM(r)}{A^3_0(r)}}\,t \right]^{2/3}\,.
\label{eq:correct}
\end{equation}
This statement is not correct. At first glance, we can already appreciate that radial and time dependences in Eq.~\eqref{eq:wrong} factorise and this is suspect for a radially inhomogeneous model, since Einstein equations are nonlinear. In fact, this factorisation does not occur in general for the solution in Eq.~\eqref{eq:correct} precisely as a consequence of the nonlinear dynamical evolution given by General Relativity in presence of radial inhomogeneties.

Moreover, we outline the following discrepancies between our model and the one discussed in \cite{Pasten:2022ggz}.
\begin{itemize}
\item 
The authors of  \cite{Pasten:2022ggz} claims that in  \cite{Cosmai:2018nvx} \ the bang time has been set to 0. This is not true since 
in Eq.~\eqref{eq:correct} the bang time is related to $2GM(r)$ and $A_0(r)$. Hence it is not a further choice on top of the setting of $A_0(r)$ and $2GM(r)$. To see that, we recall that the bang time $t_{BT}(r)$ is defined by condition $A(t_{BT}(r),r)=0$, which returns, from Eq.~\eqref{eq:correct}
\begin{equation}
t_{BT}(r)=-\frac{2}{3}\sqrt{\frac{A^3_0(r)}{2GM(r)}}\,.
\end{equation}
This shows that the solution presented in \cite{Cosmai:2018nvx} describes a inhomogeneous Big-Bang model rather than an Einstein-de Sitter one, and this is independent of the unique actual gauge fixing at our disposal, contrary to what the authors claim in their work  \cite{Pasten:2022ggz} (in \cite{Cosmai:2018nvx}, it has been settled $A_0(r)=r$).
\item As a corollary to the previous bullet point, we also have that the Hubble-Lema\^itre function reported in Eq.~(29) of  \cite{Pasten:2022ggz} is not the same as one gets from Eq.~\eqref{eq:correct} in this paper. In fact, if we define $\dot{f}\equiv\partial_t f$ and $f'\equiv\partial_r f$, we have
\begin{equation}
\frac{\dot A}{A}(t,r)=\frac{2}{3}\left[ t-t_{BT}(r) \right]^{-1}\,.
\label{eq:transverse}
\end{equation}
Hence, today ($t=0$) the inhomogeneous Hubble-Lema\^itre $H_0(r)$ function is
\begin{equation}
H_0(r)\equiv\frac{\dot A}{A}(0,r)=-\frac{2}{3\,t_{BT}(r)}\,,
\end{equation}
which is correctly positively defined, since $t_{BT}(r)<0$.
Moreover, for what concern  the Hubble functions, it is well-known \cite{Alnes:2005rw} that for the LTB models Eq.~\eqref{eq:transverse} describes the transverse Hubble-Lema\^itre flow $H_\perp(t,r)$ whereas the longitudinal expansion rate is given by
$H_\rVert(t,r)\equiv\frac{\dot A'}{A'}(t,r)$. These are equal only in the FLRW limit. From Eq.~\eqref{eq:correct}, we have then that
\begin{widetext}
\begin{equation}
\frac{\dot A'}{A'}(t,r)=\frac{2}{3}\frac{3 t_{BT}(r) A_0'(r) (t-t_{BT}(r))+ A_0(r) (3 t_{BT}(r)-2 t) t_{BT}'(r)}{3 t_{BT}(r) A_0'(r) (t-t_{BT}(r))^2+2 t A_0(r) (t_{BT}(r)-t) t_{BT}'(r)}\,,
\end{equation}
\end{widetext}
which is clearly different from $H_\rVert$ regardless of the gauge fixing for $A_0(r)$. Instead, the solution \eqref{eq:wrong} provides
\begin{equation}
\frac{\dot R}{R}=\frac{\dot R'}{R'}=\frac{2}{3t}\,,
\end{equation}
which is precisely what happens in the Einstein-de Sitter Universe.
\end{itemize}

The physical reason behind the apparent contradictory result found by Past\'en {et al.} is the following.
\begin{figure}[ht!]
\centering
\includegraphics[scale=0.51]{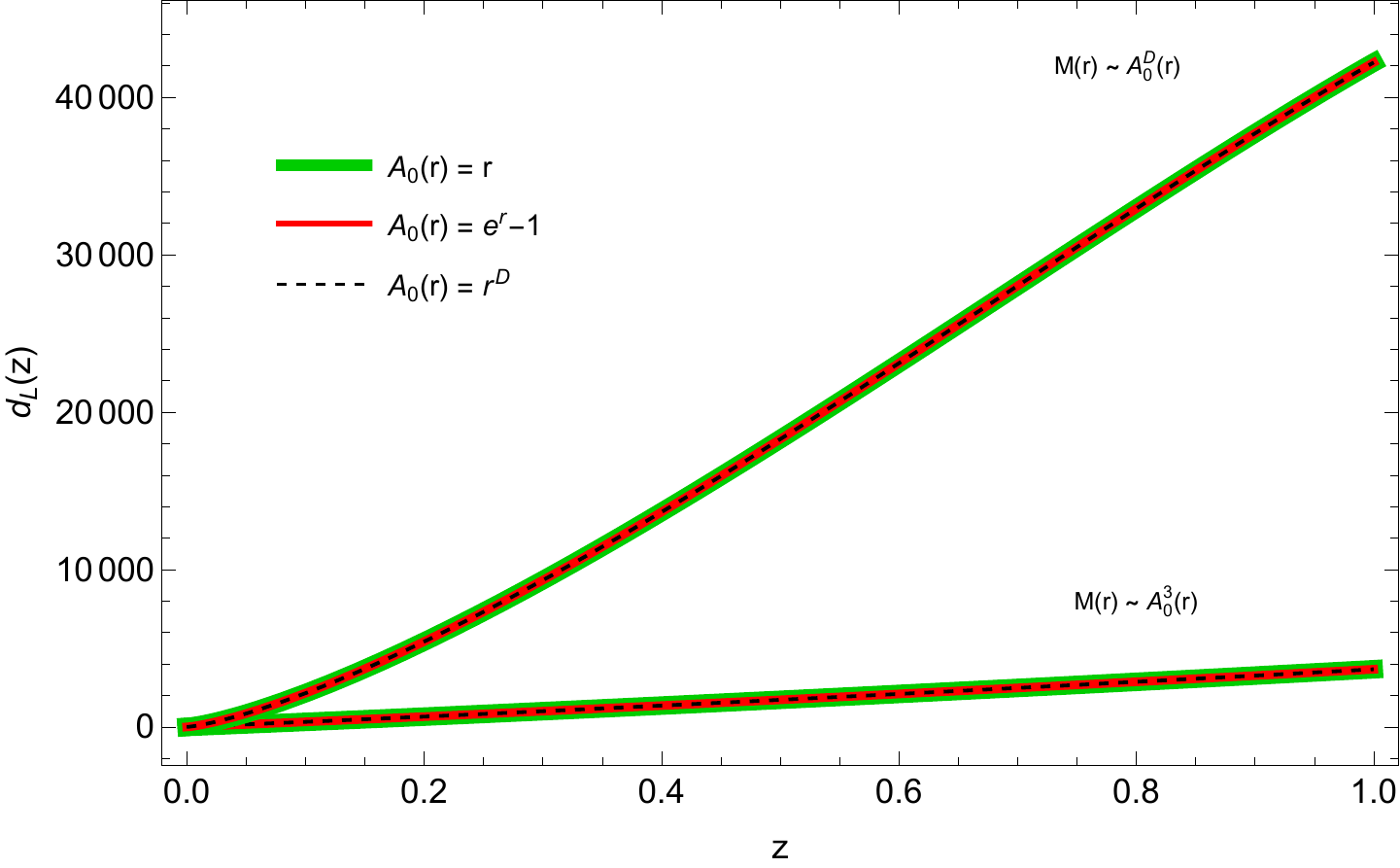}
\caption{Plot of $d_L(z)=(1+z)^2A(t(z),r(z))$ for two different cases of the Einstein-deSitter Universe ($M(r)\sim A^3_0(r)$) and for the fractal Universe ($M(r)\sim A^D_0(r)$). We have considered three different choices for the gauge fixing of $A_0(r)$ and integrated the $t(z)$ and $r(z)$ according to the Eqs.~(20)-(21) of \cite{Pasten:2022ggz} or, equivalently, Eqs.~(14) of \cite{Cosmai:2018nvx}. It is evident that the choice of $A_0(r)$ is irrelevant for the ultimate value of $d_L(z)$ and the only thing that matters is the scaling of $M(r)$ in terms of $A_0(r)$.}
\label{fig:plot}
\end{figure}
The way to connect Eq.~\eqref{eq:wrong} with Eq.~\eqref{eq:correct} is to require that $M(r)\sim A^3_0(r)$. In \cite{Pasten:2022ggz}, it is claimed that this is just a gauge fixing of the free function $A_0(r)$. On the contrary, this is precisely the physical statement that the Universe they consider is the Einstein-deSitter one. In fact, when $A_0(r)$ is not yet chosen, it plays the role of the radius, as one can directly check in the LTB line element.
Hence, stating that $M(r)\sim A^3_0(r)$ corresponds to the {\it physical assumption} that the mass in a sphere of radius $A_0(r)$ scales as $A^3_0(r)$, which is the perfectly homogeneous Universe. As a corollary, the correct way to implement a fractal distribution when $A_0(r)$ is not yet fixed is $M(r)\sim A^D_0(r)$. In \cite{Cosmai:2018nvx}, this choice reduced to $M(r)\sim r^D$ since it has been settled $A_0(r)=r$. This statement is fully supported by Fig.~\ref{fig:plot} and has a simple explanation since $d_L(z)$ is a relation between observables and then cannot depend on a coordinate redefinition.
\\
\\
Our comparison then shows that the solution provided in \cite{Cosmai:2018nvx} is compatible with the gauge freedom allowed by the coordinate redefinition and physically different from the one presented in \cite{Pasten:2022ggz}. The latter is just a coordinate rescaling allowed by the FLRW metric. We conclude then that our fractal model for the late time Universe has none of the problems highlighted in \cite{Pasten:2022ggz}.
\\
\\
{\it Acknowledgements} - GF acknowledges support by Funda\c{c}\~{a}o para a Ci\^{e}ncia e a Tecnologia (FCT) under the program {\it ``Stimulus"} with the grant no. CEECIND/04399/2017/CP1387/CT0026 and through the research project with ref. number PTDC/FIS-AST/0054/2021. GF is also member of the Gruppo Nazionale per la Fisica Matematica (GNFM) of the Istituto Nazionale di Alta Matematica (INdAM). LT is supported in part by INFN under the program
TAsP (``Theoretical Astroparticle Physics") and by the research grant number 2017W4HA7S
``NAT-NET: Neutrino and Astroparticle Theory Network", under the program PRIN 2017
funded by the Italian Ministero dell'Università\`a e della Ricerca (MUR).

\end{document}